\documentclass[acmtog,screen,nonacm]{acmart}
\acmSubmissionID{203}

\usepackage{booktabs}
\usepackage{comment}
\usepackage{multirow}
\usepackage{epsfig}
\usepackage{graphicx}
\usepackage{amsmath}
\usepackage{amssymb}
\usepackage{overpic}
\usepackage{enumitem}
\usepackage{layouts}
\usepackage{hyphenat,balance,microtype}
\usepackage[fleqn,tbtags]{mathtools}
\usepackage[capitalise]{cleveref}
\usepackage[detect-weight]{siunitx}
\usepackage{caption, subcaption, textpos}
\usepackage{colortbl}

\usepackage{algorithm}
\usepackage{algorithmic}
\usepackage{verbatim}
\usepackage{listings}
\usepackage{wrapfig}
\usepackage{makecell}
\usepackage{hyperref}
\usepackage{ulem}
\graphicspath{{./images/}}

\definecolor{Highlight}{HTML}{39b54a} 
\newcommand{\hl}[1]{\textcolor{Highlight}{\tiny{$\downarrow$}\scriptsize{#1}}}

\newcommand{\tablestyle}[2]{\setlength{\tabcolsep}{#1}
                            \renewcommand{\arraystretch}{#2}
                            \centering
                            \footnotesize}

\acmJournal{TOG}
\acmYear{2025}
\acmVolume{44}
\acmNumber{4}
\acmArticle{0}

\setcopyright{acmcopyright}

\title{OctGPT: Octree-based Multiscale Autoregressive Models for 3D Shape Generation}

\author{Si-Tong Wei}
\affiliation{
  \department{Wangxuan Institute of Computer Technology}
  \institution{Peking University}
  \country{China}
}
\email{weisitong@pku.edu.cn}

\author{Rui-Huan Wang}
\affiliation{
  \institution{Peking University}
  \country{China}
}
\email{wangps@pku.edu.cn}

\author{Chuan-Zhi Zhou}
\affiliation{
  \institution{Peking University}
  \country{China}
}
\email{2200011057@stu.pku.edu.cn}

\author{Baoquan Chen}
\affiliation{
 \institution{Peking University}
 \country{China}
}
\email{baoquan@pku.edu.cn}

\author{Peng-Shuai Wang}
\affiliation{
 \department{Wangxuan Institute of Computer Technology}
 \institution{Peking University}
 \country{China}
}
\email{wangps@hotmail.com}
\authornote{Corresponding author.}

\begin{abstract}
Autoregressive models have achieved remarkable success across various domains, yet their performance in 3D shape generation lags significantly behind that of diffusion models.
In this paper, we introduce OctGPT, a novel multiscale autoregressive model for 3D shape generation that dramatically improves the efficiency and performance of prior 3D autoregressive approaches, while rivaling or surpassing state-of-the-art diffusion models.
Our method employs a serialized octree representation to efficiently capture the hierarchical and spatial structures of 3D shapes.
Coarse geometry is encoded via octree structures, while fine-grained details are represented by binary tokens generated using a vector quantized variational autoencoder (VQVAE), transforming 3D shapes into compact \emph{multiscale binary sequences} suitable for autoregressive prediction.
To address the computational challenges of handling long sequences, we incorporate octree-based transformers enhanced with 3D rotary positional encodings, scale-specific embeddings, and token-parallel generation schemes. These innovations reduce training time by $13$ folds and generation time by $69$ folds, enabling the efficient training of high-resolution 3D shapes, e.g.,$1024^3$, on just four NVIDIA 4090 GPUs only within days.
OctGPT showcases exceptional versatility across various tasks, including text-, sketch-, and image-conditioned generation, as well as scene-level synthesis involving multiple objects.
Extensive experiments demonstrate that OctGPT accelerates convergence and improves generation quality over prior autoregressive methods, offering a new paradigm for high-quality, scalable 3D content creation.
\emph{Our code and trained models are available at \url{https://github.com/octree-nn/octgpt}.}
\end{abstract}

\begin{teaserfigure}
  \centering
  \includegraphics[width=\linewidth]{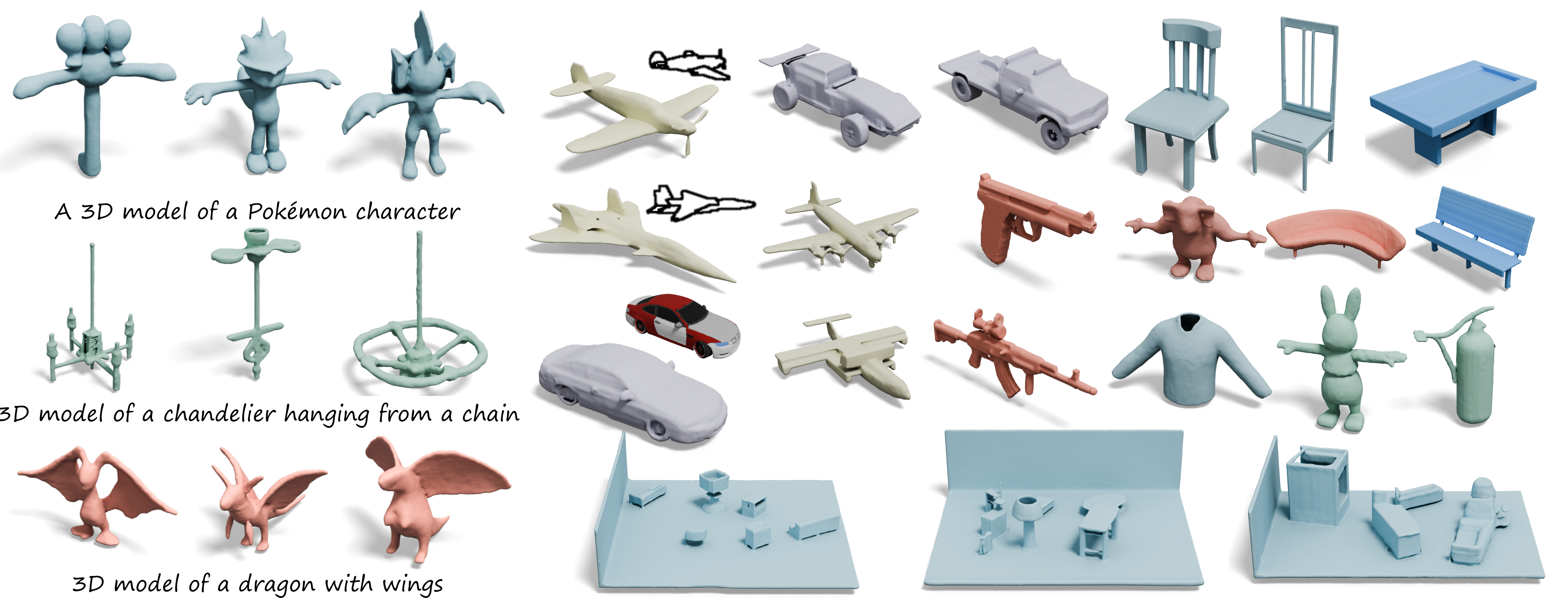}
  \caption{OctGPT generates high-quality 3D shapes in diverse scenarios, including unconditional, category-, text- and image-conditioned generation, as well as large-scale scene synthesis.}
  \label{fig:teaser}
\end{teaserfigure}



\begin{document}

\maketitle
\section{Introduction} \label{sec:intro}

3D content creation has experienced rapid advancements in recent years, driven largely by diffusion models~\cite{Zheng2023,Cheng2023,Hui2022,Li2023,Chou2023,Zhang2023a,Gupta2023,Zhang2024,Ren2024}.
Meanwhile, autoregressive models have achieved remarkable success across various domains, including large language models~\cite{Gpt3,Gpt3.5,Touvron2023}, image generation~\cite{Tian2024,Sun2024}, and large multimodal models~\cite{Gpt4,Wang2024}.
Despite these successes, the performance of autoregressive models in 3D shape generation continues to lag behind that of diffusion models.
Exploring the potential of autoregressive models in 3D content creation may unlock new possibilities for improving the quality, diversity, and scalability of generated 3D shapes, ultimately paving the way for the development of general multimodal models. \looseness=-1

Autoregressive models operate by predicting the next token based on previous tokens using transformers~\cite{Vaswani2017}.
While conceptually straightforward, applying autoregressive models to 3D shape generation presents several challenges.
The first challenge stems from the fact that autoregressive models need to define a sequential order for tokens, which does not naturally exist for 3D shapes.
Previous approaches often flatten 3D tokens into 1D sequences using rasterization orders based on 3D positions~\cite{Siddiqui2024,Chen2024,Yan2022,Zhang2022a}.
However, such approaches overlook the hierarchical structure and spatial locality inherent to 3D shapes, leading to slower convergence and suboptimal generation quality.
Another challenge lies in that 3D shapes require a large number of tokens to capture the complex geometry and topology, which makes both the training and inference processes computationally intensive.
To address this, prior works have explored compact mesh-based representations~\cite{Siddiqui2024,Chen2024,Nash2020} and low-dimensional tokenization schemes~\cite{Mittal2022,Yan2022,Zhang2022a}, reducing the token count to approximately $1k$.
Despite these efforts, these methods still face limitations in expressiveness and struggle to produce high-quality 3D shapes with fine-grained details.
\looseness=-1

In this paper, we present an efficient and scalable autoregressive model for 3D shape generation with a novel serialized octree representation, achieving quality and scalability that even surpass state-of-the-art diffusion models.
Our key observation is that octrees inherently capture the hierarchical structure of 3D shapes while providing a locality-preserving order suitable for autoregressive prediction, as octree nodes at each depth are sorted in z-order by construction~\cite{Zhou2011,Wang2023}.
We convert input shapes into octrees by recursively splitting nodes to a specified depth or resolution.
We regard the node-splitting status as a 0/1 binary signal and concatenate these signals across each octree depth from coarse to fine, forming a 1D sequence.
While the octree structure effectively captures coarse geometry, it lacks fine-grained geometric details.
To address this, we supplement the octree with additional binary tokens defined at the finest octree nodes, generated by a vector quantized variational autoencoder (VQVAE)~\cite{Van2017} that is built on octree-based neural networks~\cite{Wang2017,Wang2022}, using binary spherical quantization~\cite{Zhao2024}.
These binary tokens are concatenated with the binary splitting signals to form the final input sequence for the autoregressive model.
The autoregressive model is trained to predict the binary sequence through a series of binary classification tasks.
During inference, the predictions are assembled to reconstruct the octree structure, from coarse to fine, along with the binary tokens at the finest octree nodes, which are decoded into continuous signed distance fields (SDFs) by the decoder of the VQVAE.
The SDFs are subsequently converted to meshes using the marching cubes algorithm~\cite{Lorensen1987}  as the final output. \looseness=-1

In contrast to previous autoregressive models that directly predict 3D coordinates~\cite{Siddiqui2024,Chen2024,Yan2022,Zhang2022a}, our approach decomposes the generation task into a series of simpler binary classification tasks.
This methodology is inspired by the chain-of-thought paradigm~\cite{Wei2022}, which improves the reasoning capabilities of large language models by breaking down complex reasoning into a sequence of intermediate steps.
By adopting this approach, we achieve significantly faster convergence and higher generation quality compared to direct coordinate prediction.
For instance, in one of our experiments, our method generated high-quality 3D shapes after just 10 epochs of training, whereas the coordinate prediction approach failed to produce satisfactory results even after 100 epochs. \looseness=-1

The token length of our serialized octree representation can exceed $50k$, posing a challenge for naive autoregressive models, which have quadratic time complexity.
To circumvent this issue, we adapt octree-based transformers~\cite{Wang2023} with several improvements that reduce time complexity to linear, achieving a speedup of $13$ folds.
As a result,  our model can to be trained on four NVIDIA 4090 GPUs with 24GB of memory.
We further incorporate the parallel token prediction scheme proposed by~\cite{Li2024}, accelerating the generation process by $69$ folds and enabling the generation of high-resolution 3D shapes with resolution $1024^3$ in under 30 seconds on a single NVIDIA 4090 GPU.
Additionally, we extend rotary positional encodings~\cite{Wang2023} to 3D and introduce scale-specific positional encodings for each token, enhancing the model's ability to distinguish multiscale binary signals in the serialized octree representation.
As our training and prediction process is in a GPT-like style~\cite{Gpt3}, we name our model OctGPT. \looseness=-1

We evaluate the effectiveness of our OctGPT across various 3D shape generation tasks, including shape generation on ShapeNet \cite{Chang2015} and Objaverse~\cite{Deitke2023}, text- or image-conditioned generation, and large-scale scene generation.
The generation results showcased in \cref{fig:teaser} highlight the high-quality 3D shapes at a resolution of $1024^3$ produced in diverse scenarios.
Quantitative evaluations demonstrate that OctGPT significantly outperforms previous 3D autoregressive models~\cite{Zhang2022a,Mittal2022,Chen2024} in terms of generation quality, diversity, and scalability, and even surpasses state-of-the-art 3D diffusion models in certain cases.
OctGPT offers a new paradigm for 3D content creation, distinct from diffusion models.
We believe it can inspire further exploration within the 3D generation community and promote the development of multimodal models through alignment or fine-tuning from large language and image models~\cite{Liu2024}.
In summary, our main contributions are as follows:
\begin{itemize}[leftmargin=*,itemsep=2pt]
  \item[-] We propose a 3D autoregressive model trained on serialized multiscale binary tokens induced by octrees for 3D shape generation, significantly improving the efficiency and scalability of 3D autoregressive models compared to previous works.
  \item[-] We extend octree-based transformers to handle multiscale octree nodes and adapt a parallel token prediction scheme, accelerating training speed by $13$ folds and generation speed by $69$ folds.
  \item[-] We demonstrate that autoregressive models can be effectively applied to 3D shape and scene generation, achieving competitive results with state-of-the-art diffusion models, while requiring only 3 days of training on 4 Nvidia 4090 GPUs.
\end{itemize}

\section{Related Works}

\paragraph{3D Generation}
Early works in 3D shape generation utilized generative adversarial networks~\cite{Goodfellow2016a} to synthesize 3D shapes from random noise~\cite{Wu2016,Chen2019,Zheng2022}.
More recently, the focus has shifted toward adapting diffusion models~\cite{Ho2020} for 3D generation across various 3D representations, including voxels~\cite{Cheng2023,Li2023,Hui2022,Chou2023,Shim2023}, point clouds~\cite{Zeng2022,Lou2021,Nichol2022}, triplanes~\cite{Shue2023,Gupta2023}, implicit shape functions~\cite{Zhang2023a,Jun2023,Erkoc2023,Zhang2024}, Gaussian splattings~\cite{Roessle2024}, and sparse voxel structures~\cite{Zheng2023,Ren2024,Xiong2024,Xiang2024,Liu2023,xiong2024b}.
While these methods have achieved remarkable results, their generation paradigms differ fundamentally from those of autoregressive models~\cite{Gpt4,Touvron2023,Wang2024}.
In this paper, we explore the potential of autoregressive models for 3D shape generation, with the aim of laying a foundation for general multimodal models capable of simultaneously generating 3D shapes, images, and text. \looseness=-1

\paragraph{3D Autoregressive Models}
Inspired by the success of autoregressive models in text and image generation, several pioneering works have begun exploring their application to 3D shape generation.
A few methods leverage autoregressive models to generate 3D meshes~\cite{Nash2020,Siddiqui2024,Chen2024,Wang2024a,Chen2024a,Weng2024,Hao2024}, mainly by predicting vertices and faces in a predefined rasterization order.
There are also works that autoregressively generate 3D shapes in point clouds~\cite{Sun2018,Cheng2022a}.
Following trends in image generation, an intuitive extension involves training a VQVAE~\cite{Van2017} to compress 3D shapes into low-dimensional latent spaces~\cite{Zhang2022a,Yan2022,Mittal2022,Qian2024,Cheng2022a} before training autoregressive models to improve efficiency.
However, these methods face challenges in capturing complex geometric details due to the long sequence lengths required to encode intricate textures and structures of 3D shapes.
Additionally, their serialization methods often disregard the hierarchical structure and spatial locality inherent to 3D shapes, leading to slower training convergence.
Autoregressive models have also been explored for scene layout generation~\cite{Wang2021d,Paschalidou2021}.
\emph{Concurrently}, recent works have extended the next-scale prediction paradigm, initially introduced for image generation~\cite{Tian2024}, to 3D shape generation by constructing multiscale latent spaces~\cite{Zhang2024a,Chen2024b,Zhang2024b}.
Our work addresses limitations  of previous works by proposing a novel autoregressive model that generates 3D shapes in a novel binarized octree representation and introduces efficient transformer architectures to capture long-range dependencies in token sequences. \looseness=-1

The work most closely related to our OctGPT is the Octree Transformer proposed by~\cite{Ibing2023}.
However, while Octree Transformers primarily focus on compressing octree sequence lengths for efficient training, our method introduces novel transformer architectures and generation schemes to enhance efficiency.
Additionally, Octree Transformers predict the coordinates of octree nodes directly, whereas our approach predicts the binary splitting status of octree nodes, significantly improving the quality of the generated shapes.
Moreover, Octree Transformers are limited to generating sparse voxels, whereas our method produces continuous implicit fields, which can be converted into high-resolution meshes. \looseness=-1

\paragraph{3D Transformers}
Autoregressive models primarily rely on transformers~\cite{Vaswani2017} to learn dependencies within token sequences.
Transformers have been adapted to 3D data by treating each point or patch as a token and applying attention mechanisms globally across all tokens~\cite{Guo2021,Yu2022,Pang2022}.
However, these methods are limited to small-scale point clouds with only a few thousand tokens due to the huge computational and memory costs of global attention mechanisms.
To address these issues, several works have proposed to constrain attention modules to operate within local neighborhoods~\cite{Zhao2021,Wu2022} or non-overlapping windows of point clouds~\cite{Fan2022,Sun2022,Lai2022,Yang2024}.
More recently, the OctFormer~\cite{Wang2023} and its variants~\cite{Wu2024} introduced a novel approach by serializing point tokens along z-order or Hilbert curves and partitioning the resulting token sequences into windows with equal token counts before applying attention mechanisms.
This design achieves state-of-the-art performance on large-scale 3D understanding benchmarks.
Our work builds on the principles of OctFormer, but we shift the focus from 3D understanding tasks to autoregressive 3D shape generation.
Furthermore, we extend OctFormer to handle multiscale token sequences by incorporating scale-specific embeddings into the transformer architecture.
\looseness=-1

\section{Method} \label{sec:method}

\begin{figure*}[t]
  \centering
  \includegraphics[width=0.95\linewidth]{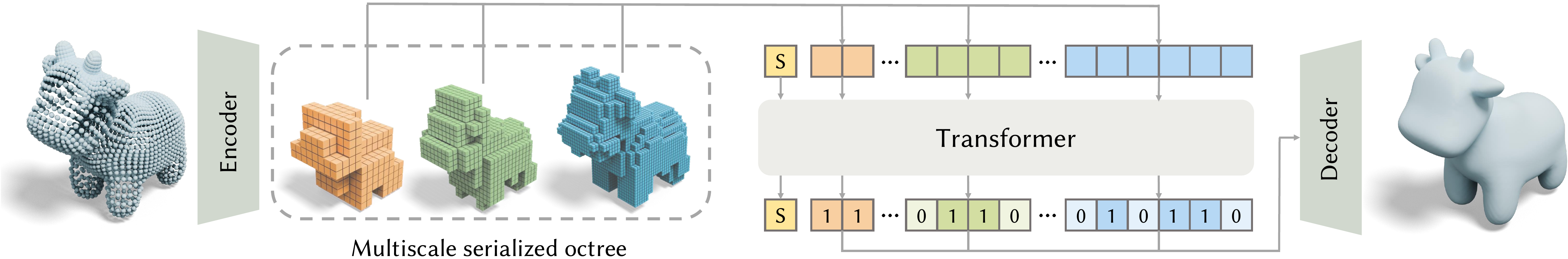}

  \caption{Overview.
  3D shapes are encoded as multiscale serialized octrees, where coarse structures are represented by multiscale binary splitting signals derived from the octree hierarchy, and fine-grained details are captured by binarized latent codes from an octree-based VQVAE.
  These binary tokens, along with teacher-forcing masks, are fed into a transformer for autoregressive training.
  During inference, the transformer progressively predicts the token sequence to reconstruct the octree and latent codes, generating 3D shapes from coarse to fine. The sequence is decoded by the VQVAE to produce the final 3D shape.
   }
  \label{fig:overview}
\end{figure*}

Our goal is to harness autoregressive models for 3D shape generation.
This involves two key components: a serialized representation of 3D shapes and an autoregressive model.
The overview of our model is presented in \cref{fig:overview}.
We first introduce a novel serialized octree representation, produced by an octree-based VQVAE~\cite{Van2017}, which encodes 3D shapes into multiscale binary token sequences.
The details of this representation are elaborated in \cref{sec:vqvae}.
Next, we train transformers to autoregressively predict these sequences, enabling the generation of 3D shapes from coarse to fine.
To enhance efficiency and performance, we incorporate octree-based attention mechanisms~\cite{Wang2023}, multi-token prediction strategies~\cite{Li2024}, and customized 3D positional encodings, achieving substantial improvements in training and generation performance, as described in \cref{sec:ar}.

\subsection{Serialized Octree Representation} \label{sec:vqvae}

Our representation consists of two components: an octree and quantized binary codes produced by an octree-based VQVAE.
The octree captures the coarse geometric structure of the shape, while the VQVAE encodes fine-grained details and is responsible for reconstructing the continuous signed distance fields (SDFs).
Both the octree structure and the quantized codes are serialized into a binary sequence, serving as input to the subsequent autoregressive model.

\subsubsection{Serialized Octree}
\begin{figure}[t]
  \centering
  \includegraphics[width=0.95\linewidth]{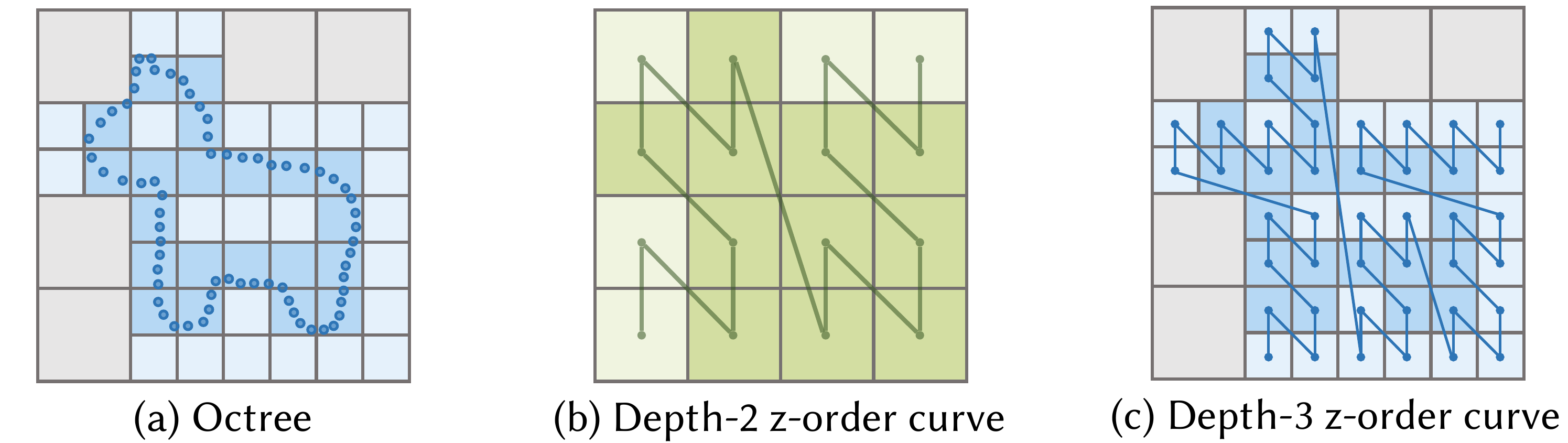}

  \caption{Octree and z-order curves. 2D images are used for clearer illustration.
  (a): The input point cloud with its corresponding octree.
  Node statuses are color-coded: darker colors represent nodes containing points, lighter colors indicate empty nodes, and gray denotes non-existing nodes at the given depth.
  (b) \& (c): z-order curves at octree depths 2 and 3, respectively.}
  \label{fig:octree}

\end{figure}

Given a 3D shape, we first rescale it to fit within a unit cube and construct an octree by recursively subdividing non-empty voxels until a specified depth is reached~\cite{Zhou2011,Wang2023}.
Additionally, we enforce the first three levels of the octree to be fully populated, which covers the entire volume, including any disconnected components.
The octree structure is determined by the split status at each node, with nodes sorted in z-order according to shuffled keys~\cite{Zhou2011}.
Notably, spatial locality is well preserved in the octree due to the z-order, \emph{i.e.}, octree nodes that are spatially close are likely to be adjacent in the z-order curve. An example is shown in \cref{fig:octree}. \looseness=-1

Denote the maximum depth of the octree as $D$, and the node split status at depth $d$ as $o_i^d \in \{0, 1\}$, where $i$ is the node index in the z-order curve and $ 0/1 $ indicate no split/split, respectively.
The octree nodes at depth $d$ are concatenated into a binary sequence $ O^d = (o_1^d, o_2^d, \dots, o_{n_d}^d) $, where $n_d$ is the number of nodes at depth $d$.
At depth 3, the octree is fully populated, resulting in 512 nodes.
Starting from depth 3, the binary sequences from each level up to depth $(D - 1)$ are concatenated, forming a multiscale binary sequence $\mathcal{O} = (O^3, O^4, \dots, O^{D-1}) $.
Inversely, this sequence $ \mathcal{O} $ can be used to reconstruct the octree structure, effectively capturing the coarse geometry  of a 3D shape.

\subsubsection{Octree-based VQVAE}
\begin{figure}[t]
  \includegraphics[width=\linewidth]{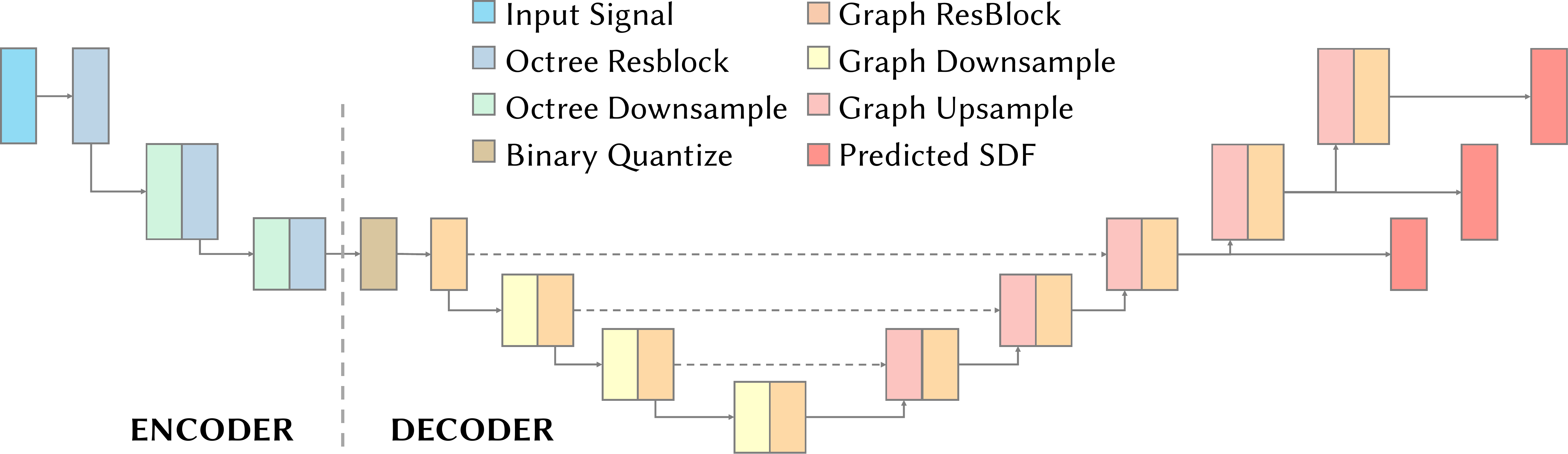}
  \caption{The architecture of Octree-based VQ-VAE. The encoder compresses the input octree signals with octree-based residual blocks and reduce the depth of the octree by 2. The features are then quantized into binary tokens and fed into the decoder. The decoder builds a dual octree graph and applies graph convolution to predict SDFs for shape reconstruction. }
  \label{fig:vqvae}
\end{figure}

Complementing the octree, we introduce an octree-based VQVAE for efficient tokenization and reconstruction of SDFs and geometric details of 3D shapes.
The VQVAE generates \emph{quantized binary codes} at the finest level of the octree.
The detailed architecture of the VQVAE is illustrated in \cref{fig:vqvae}.
Our VQVAE model employs an asymmetric encoder-decoder architecture, with a particular emphasis on enhancing the decoder's ability to achieve high-fidelity surface reconstruction.

The encoder is based on octree-based CNNs (O-CNN)~\cite{Wang2017}.
Starting with an octree constructed from a 3D mesh or point cloud, the encoder efficiently compresses the input octree into feature representations at the leaf nodes, reducing the depth of the octree by 2.
These features are then quantized into binary tokens using the Binary Spherical Quantization (BSQ) method~\cite{Zhao2024}.
Specifically, the feature vector $z_i \in \mathbb{R}^d$ at the $i^{th}$ leaf node is quantized into binary tokens $q_i$ as follows: $q_i = \operatorname{sign} \left( \frac{z_i}{\|z_i\|} \right)$.
The BSQ eliminates the need for codebooks, as required in conventional VQ-VAEs, thereby simplifying implementation while maintaining comparable reconstruction quality.
Following~\cite{Zhao2024}, we also employ a loss function, denoted as $\mathcal{L}_{vq}$, to encourage the binary tokens to be uniformly distributed. \looseness=-1

Subsequently, the decoder employs the dual octree graph networks~\cite{Wang2022} to decode the binary tokens $q_i$ into local SDFs, which are then integrated into a global SDF using the multi-level Partition-of-Unity method~\cite{Ohtake2003,Wang2022}.
This process ensures a coherent and detailed reconstruction of the original shape.
The loss for SDF reconstruction is defined as:
\begin{equation}
  \mathcal{L}_{sdf} = \frac{1}{N_\mathcal{P}}  \sum_{x \in \mathcal{P}} \left( \lambda \| S(x) - D(x)  \|_2^2 + \| \nabla S(x) -  \nabla D(x) \|_2^2 \right),
  \label{equ:sdf}
\end{equation}
where $S(x)$ and $D(x)$ are the predicted and ground-truth SDFs, respectively, $\lambda$ is a weighting factor, $\mathcal{P}$ is the set of sampled points, and $N_\mathcal{P}$ is the number of sampled points.
This loss encourages the decoder to reconstruct the continuous SDFs accurately.
Since the encoder reduces the depth of the octree, the decoder is designed to recover the original octree depth with an octree splitting loss $\mathcal{L}_{octree}$, following~\cite{Wang2022,Wang2018a}.

In total, the loss of the VQVAE is defined as:
\begin{equation}
  \mathcal{L} = \mathcal{L}_{vq} + \mathcal{L}_{sdf} + \mathcal{L}_{octree}.
  \label{equ:vqvae}
\end{equation}
The binary tokens $q_i$ produced by the encoder are also serialized according to the z-order curve, forming the sequence $\mathcal{Q}$, which is concatenated with $\mathcal{O}$ to create our final serialized octree representation. \looseness=-1

\paragraph{Remarks}
Our serialized octree representation consists entirely of binary tokens, which can be easily processed and predicted by autoregressive models.
Furthermore, our representation captures multiscale information of the 3D shape and preserves the spatial locality of octree nodes. 
These properties make it more suitable for autoregressive prediction compared to the commonly used rasterization order in previous works~\cite{Zhang2022a,Siddiqui2024,Yan2022}, as examined in our experiments in \cref{sec:ablation}.

\subsection{Multiscale Autoregressive Models} \label{sec:ar}
Using our serialized octree representation, we can train autoregressive models to predict it, enabling the progressive generation of 3D shapes from coarse to fine.
While directly applying standard autoregressive models~\cite{Touvron2023,Gpt3} to this representation is feasible, it is computationally inefficient and often results in suboptimal performance.
To overcome these challenges, we introduce three key enhancements, including an efficient transformer architecture, customized positional encoding, and multi-token generation strategies, significantly improving the training and generation efficiency.

\subsubsection{Efficient Transformer Architecture}
\begin{figure}[t]
  \centering
  \includegraphics[width=0.95\linewidth]{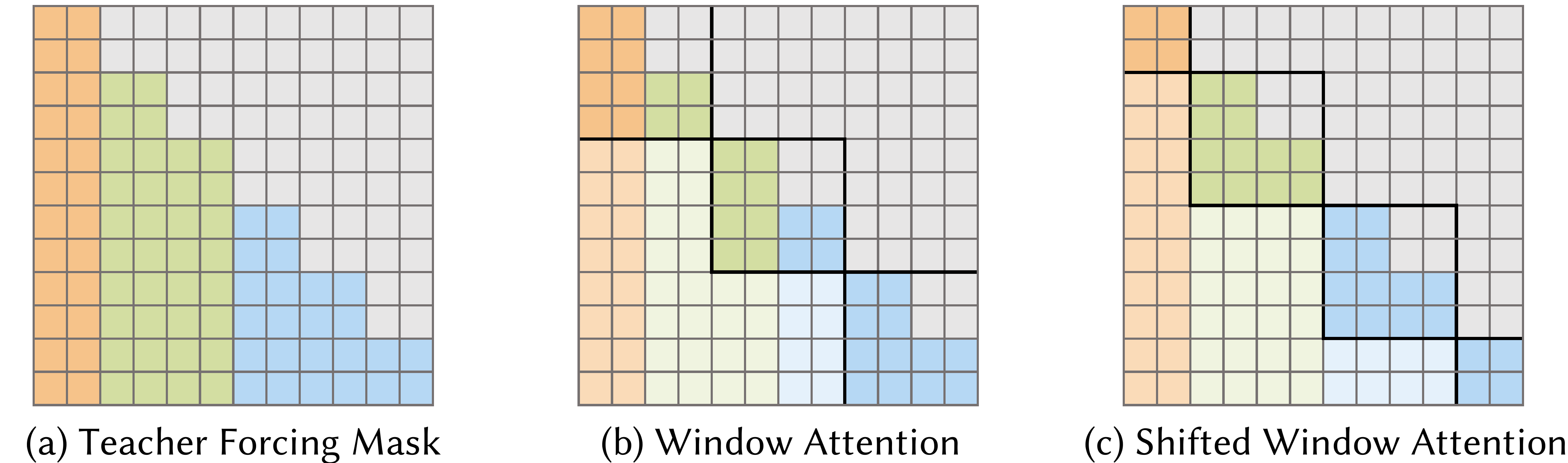}
  \caption{Multiscale Autoregressive Models.
  (a) Our model predicts multiple tokens autoregressively according to the depth-wise teacher-forcing mask.
  Tokens at different scales are represented in distinct colors, while masks are depicted in gray.
  (b) Octree-based Window attention is adopted for cross-scale communication and improved computational efficiency.
  (c) Shifted window attention allows for interactions across different windows.
  }

  \label{fig:mask}
\end{figure}

Our transformer architecture consists of a stack of attention blocks~\cite{Vaswani2017}, each containing a multi-head self-attention module and a feed-forward network with layer normalization and residual connections.
Given that token sequences can exceed $50k$ in length, computing self-attention over all tokens is computationally prohibitive.
To address this, we adopt octree-based attentions (OctFormer)~\cite{Wang2023}, which divide tokens into fixed-size windows for efficient self-attention computation.
Furthermore, we alternate between \emph{dilated} octree attention and \emph{shifted} window attention~\cite{Yang2024}, enabling interactions across different windows.
The illustration of octree-based attention and shifted window attention is shown in \cref{fig:mask}-(b)\&(c).
Unlike OctFormer, which restricts tokens to a single octree depth, our design accommodates tokens from nodes at varying depths, allowing comprehensive interactions among all tokens.
This approach captures both local and global dependencies, enhancing the model's ability to represent 3D shapes.

\subsubsection{Positional Encoding}
As transformers are permutationally invariant, positional information must be incorporated into the model.
Each token in the sequence corresponds to an octree node with a 3D position within the unit cube and an associated depth.
To encode positional information, we propose RoPE3D that extends rotary positional encoding (RoPE)~\cite{Su2024} to 3D space, inspired by the RoPE2D design~\cite{Heo2025}.
Additionally, we introduce learnable scale embeddings to differentiate tokens at different octree depths.
By integrating 3D positional encoding with scale embeddings, our transformer can effectively distinguish tokens based on their spatial positions and depths, which is crucial for accurate autoregressive prediction of 3D shapes.
The effectiveness of our positional encoding is demonstrated through experiments presented in \cref{sec:ablation}.

\subsubsection{Multiple Token Generation}
Standard autoregressive models predict one token at a time, which can be inefficient for generating 3D shapes due to the large number of tokens involved.
To address this limitation, we adopt the multi-token generation strategy proposed in masked autoregressive models (MAR)~\cite{Li2024}, which enables the parallel prediction of multiple tokens.
This approach significantly reduces the number of forward passes required for 3D shape generation, thereby improving efficiency.

MAR operates by randomly permuting tokens in the sequence and masking a portion of them, which are then predicted by the model in an autoregressive manner.
However, directly applying MAR to our sequence can lead to dependency issues since our sequence comprises tokens from multiple octree depths, where tokens at deeper octree layers depend on those from shallower layers.
To address this challenge, we introduce a depth-wise teacher-forcing mask, as illustrated in \cref{fig:mask}-(a).
Instead of permuting the whole sequence, we permute tokens within each depth level, ensuring that tokens at higher depths can access information from lower-depth tokens while preventing information leakage from higher-depth tokens.
By doing so, it preserves the hierarchical dependencies in the sequence.  \looseness=-1

During inference, token prediction begins at depth 3 and proceeds sequentially to the maximum octree depth $D$.
Tokens from depths 3 to ($D-1$) are interpreted as split signals, progressively generating coarse voxel grids and refining them into deeper octrees.
Tokens at depth $D$ are decoded as binary quantized codes, capturing the fine details of the 3D shape.
Both split signals and quantized codes are binary tokens and are predicted using binary classifiers attached on the transformer.
The predicted sequence is then fed into the VQVAE decoder to reconstruct the final 3D shape.

\section{Experiments}  \label{sec:exp}

In this section, we evaluate the performance and efficiency of our OctGPT on 3D shape generation tasks in \cref{sec:3dgen}, conduct ablation studies in \cref{sec:ablation}, and demonstrate the versatility of our model through various applications in \cref{sec:applications}.
All experiments are performed using \emph{NVIDIA 4090 GPUs} with 24GB of memory.

\subsection{3D Shape Generation} \label{sec:3dgen}
In this section, we evaluate the performance of our model and compare it with state-of-the-art methods on the ShapeNet~\cite{Chang2015} and Objaverse~\cite{Deitke2023} datasets. The datasets and implementation details are provided in the supplementary material. \looseness=-1

\begin{figure}
  \centering
  \includegraphics[width=0.99\linewidth]{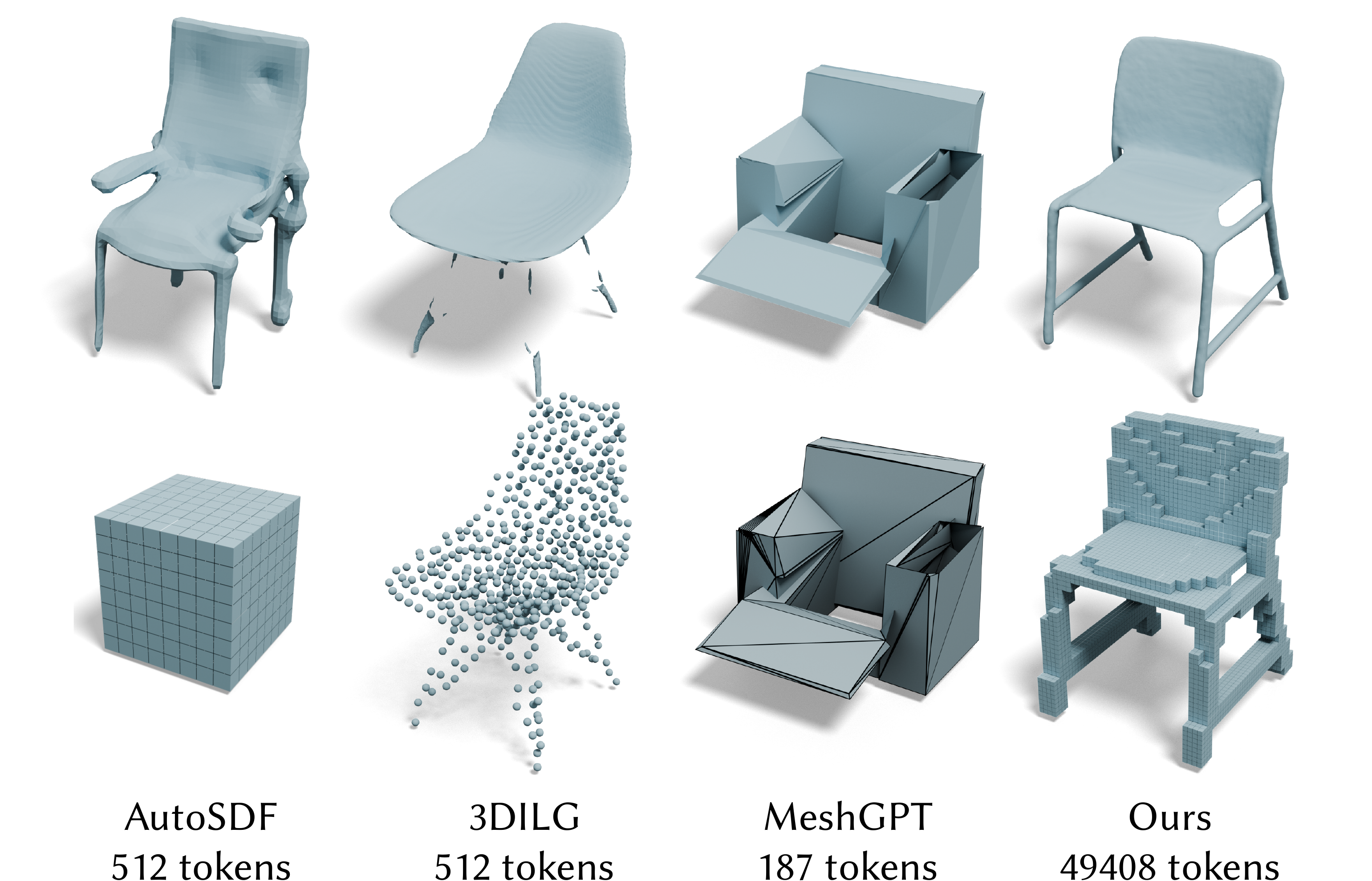}
  \caption{Comparision with state-of-the-art 3D autoregressive models. Experiments are conducted on the chair category. Top: the generated shapes. Bottom: the corresponding token. }
  \label{fig:autoregressive}
\end{figure}

\paragraph{Comparisons}
We conduct comparisons with state-of-the-art 3D generation methods, including IM-GAN~\cite{Chen2019}, SDF-StyleGAN~\cite{Zheng2022}, Wavelet-Diffusion~\cite{Hui2022}, MeshDiffusion~\cite{Liu2023c}, SPAGHETTI~\cite{Hertz2022}, LAS-Diffusion~\cite{Zheng2023}, 3DILG~\cite{Zhang2022a}, and 3DShape2VecSet~\cite{Zhang2023a}.
Among them, IM-GAN and SDF-StyleGAN are GAN-based methods, Wavelet-Diffusion, MeshDiffusion, SPAGHETTI, and LAS-Diffusion are diffusion-based methods, while AutoSDF, 3DILG, and MeshGPT are autoregressive methods, which are most closely related to our approach.
We did not include comparisons with MeshAnything~\cite{Chen2024}, MeshAnything-v2~\cite{Chen2024a}, or Edgerunner~\cite{tang2024edgerunner}, as these methods mainly focus on artistic mesh generation conditioned on point clouds, dense meshes, or images. \looseness=-1

\begin{figure*}
    \centering
    \includegraphics[width=0.95\linewidth]{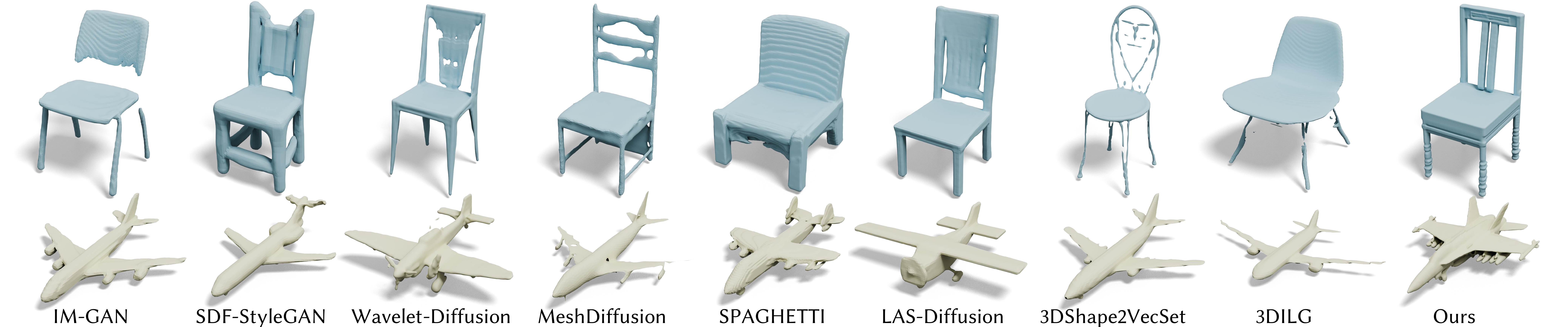}
    \caption{Qualitative comparison results. We show the generated shapes from our OctGPT and other SOTA methods on airplane and chair category.}
    \label{fig:qualitative}
\end{figure*}

The quantitative FID results are presented in \cref{tab:quantitative}.
Note that Wavelet-Diffusion was trained on only three categories, SPAGHETTI and MeshGPT were trained on two categories only, and the training datasets for MeshGPT, 3DILG and 3DShape2VecSet differ slightly from those of the other methods.
The upper part of the table shows the results for training on each category separately, while the lower part presents the results for training on all categories together.
We perform category-conditional generation when trained on all categories together.
Our OctGPT consistently outperforms other methods across most categories, even surpassing state-of-the-art diffusion-based methods like OctFusion and XCube in the \textit{Car} and \textit{Rifle} categories.
Qualitative results are shown in \cref{fig:qualitative}, where our method generates more detailed shapes than the competing approaches.
Note that diffusion-based methods~\cite{Zheng2023,Ren2024} often require multiple cascaded stages or auxiliary models, resulting in increased implementation complexity and potential error accumulation.
Moreover, diffusion models fundamentally differ from autoregressive frameworks in their generative mechanisms, which may limit their ability to achieve seamless multimodal integration. \looseness=-1

\begin{table}[!t]
  \caption{The quantitative comparison of \emph{shading-image-based FID}.
  The upper part shows results for training on each category separately, while the lower part shows results for training on all categories together.
  Shaded rows indicate autoregressive models, and green numbers highlight the improvement of our OctGPT in FID score over the previous best-performing autoregressive models. On average, OctGPT achieves the best performance.
  }
  \label{tab:quantitative}
  \tablestyle{2pt}{1.2}
  \begin{tabular}{llllll}
    \toprule
    \textbf{Method} & \textbf{Chair} & \textbf{Airplane} & \textbf{Car} & \textbf{Table} &  \textbf{Rifle} \\
    \midrule
    IM-GAN & 63.42 & 74.57 & 141.2 & 51.70 & 103.3 \\
    SDF-StyleGAN & 36.48 & 65.77 & 97.99 & 39.03 & 64.86 \\
    Wavelet-Diffusion & 28.64 & 35.05 & N/A & 30.27 & N/A \\
    MeshDiffusion & 49.01 & 97.81 & 156.21 & 49.71 & 87.96 \\
    SPAGHETTI & 65.26 & 59.21 & N/A & N/A & N/A \\
    LAS-Diffusion & 20.45 & 32.71 & 80.55 & 17.25 & 44.93 \\
    XCube & 18.07 & \textbf{19.08} & 80.00 & N/A & N/A \\
    OctFusion & \textbf{16.15} & 24.29 & 78.00 & \textbf{17.19} & 30.56  \\
    \rowcolor{gray!10} MeshGPT & 37.05 & N/A & N/A & 25.25 & N/A \\
    \rowcolor{gray!10} OctGPT  & 31.05 \hl{6.00} & 27.47 & \textbf{64.45} & 19.64 \hl{5.61} & \textbf{21.91} \\
    \midrule
    3DShape2VecSet & 21.21 & 46.27 & 110.12 & 25.15 & 54.20 \\
    LAS-Diffusion & 21.55 & 43.08 & 86.34 & \textbf{17.41} & 70.39  \\
    OctFusion & \textbf{19.63} & 30.92 & 80.97 & 17.49 & 28.59 \\
    \rowcolor{gray!10} 3DILG  & 31.64 & 54.38 & 164.15 & 54.13 & 77.74 \\
    \rowcolor{gray!10} OctGPT & 28.28 \hl{3.36} & \textbf{29.27} \hl{25.11} & \textbf{62.40}  \hl{101.75} & 20.64 \hl{33.49} & \textbf{27.21} \hl{50.53} \\
    \bottomrule
  \end{tabular}
\end{table}

\begin{figure*}
    \centering
    \includegraphics[width=0.95\linewidth]{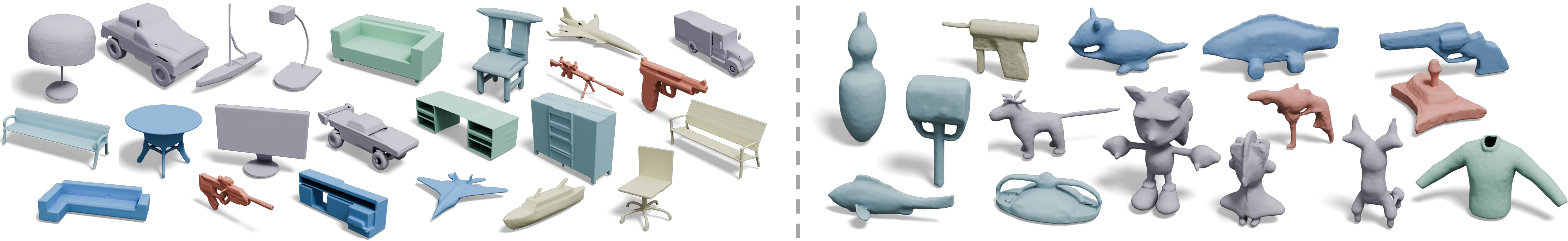}
    \caption{Unconditional and category-conditional generation results on ShapeNet dataset(left). Unconditional generation results on Objaverse dataset(right).}
    \label{fig:unconditional}
\end{figure*}

\begin{figure*}
    \centering
    \includegraphics[width=0.95\linewidth]{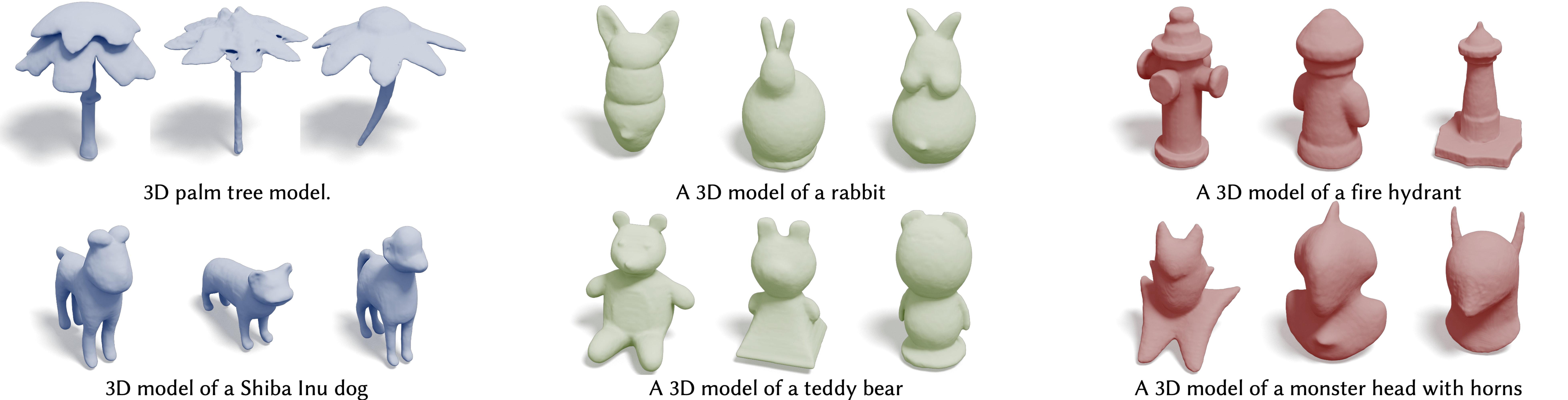}
    \caption{Text conditional generation results on Objaverse dataset.}
    \label{fig:objaverse-text}
\end{figure*}

\begin{figure*}
    \centering
    \includegraphics[width=0.95\linewidth]{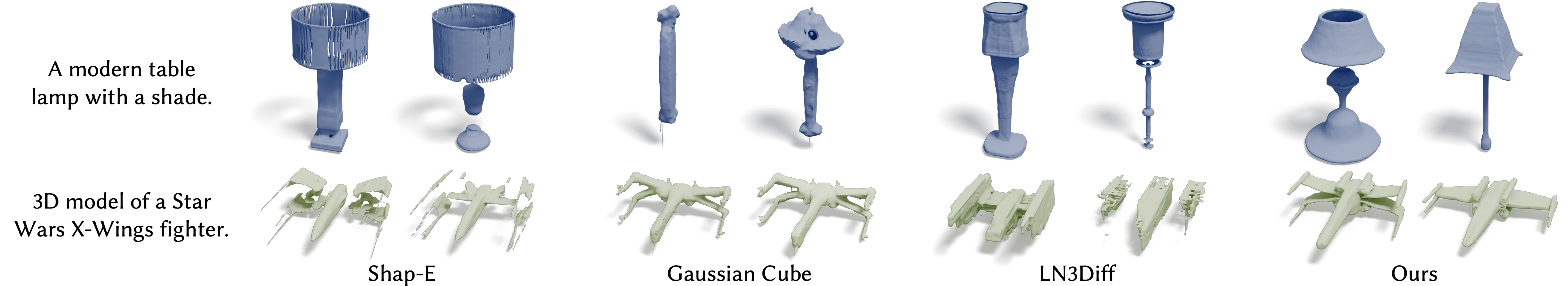}
    \caption{Comparison of text conditional generation results with other SOTA methods on Objaverse dataset.}
    \label{fig:objaverse-compare}
\end{figure*}

\begin{figure*}
    \centering
    \includegraphics[width=0.95\linewidth]{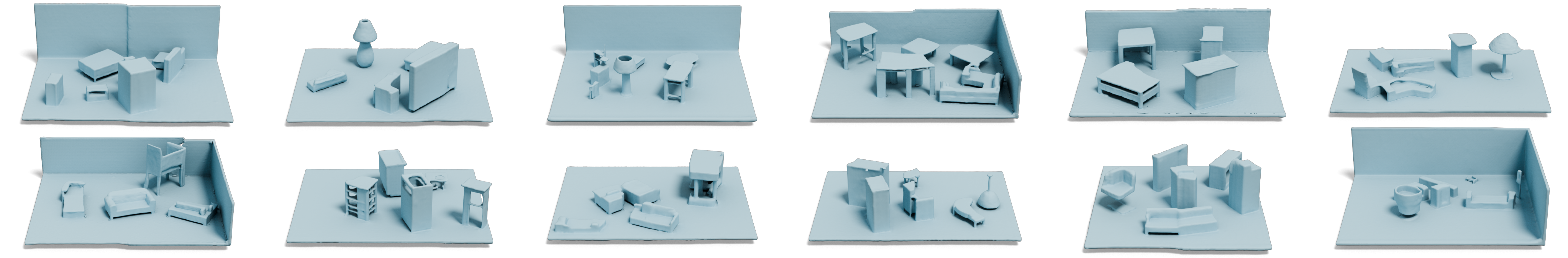}
    \caption{Scene-level generation results. Experiments are conducted on Synthetic Rooms dataset.
    }
    \label{fig:scene}
\end{figure*}

Compared to previous autoregressive methods such as AutoSDF, 3DILG and MeshGPT, OctGPT achieves a significant improvement in FID performance.
Specifically, OctGPT achieves an average improvement of 42.84 in FID score over 3DILG, which is the previous best-performing autoregressive model for 3D shape generation.
We visualize the token sequences of these autoregressive methods in \cref{fig:autoregressive}.
OctGPT is capable of processing significantly longer sequences, allowing it to capture finer details and generate higher-quality shapes.
For efficiency, OctGPT achieves a $34\times$ speedup over 3DILG at a sequence length of 80K, while AutoSDF and MeshGPT run out of memory at this sequence length, demonstrating OctGPT’s superior memory and computational efficiency.
Moreover, OctGPT uses fewer parameters, requires fewer training epochs, and consumes less GPU memory, highlighting its faster convergence and significantly reduced computational demands.
Detailed comparisons are provided in the supplementary material. \looseness=-1

\paragraph{More Generation Results}
We also train our model on 13 categories from the ShapeNet and the Objaverse dataset to demonstrate its scalability to a larger number of categories.
The results are presented in \cref{fig:unconditional}.
Our model generates high-quality and diverse shapes across different categories.
The generated shapes exhibit rich details, and the overall quality remains consistent with the results obtained from the five categories of ShapeNet. \looseness=-1

\subsection{Ablation Study and Discussions} \label{sec:ablation}

In this section, we conduct ablation studies to analyze the impact of key design choices in our model.
We conduct experiments on the airplane category from ShapeNet. \looseness=-1

\begin{figure}
  \centering
  \includegraphics[width=0.99\linewidth]{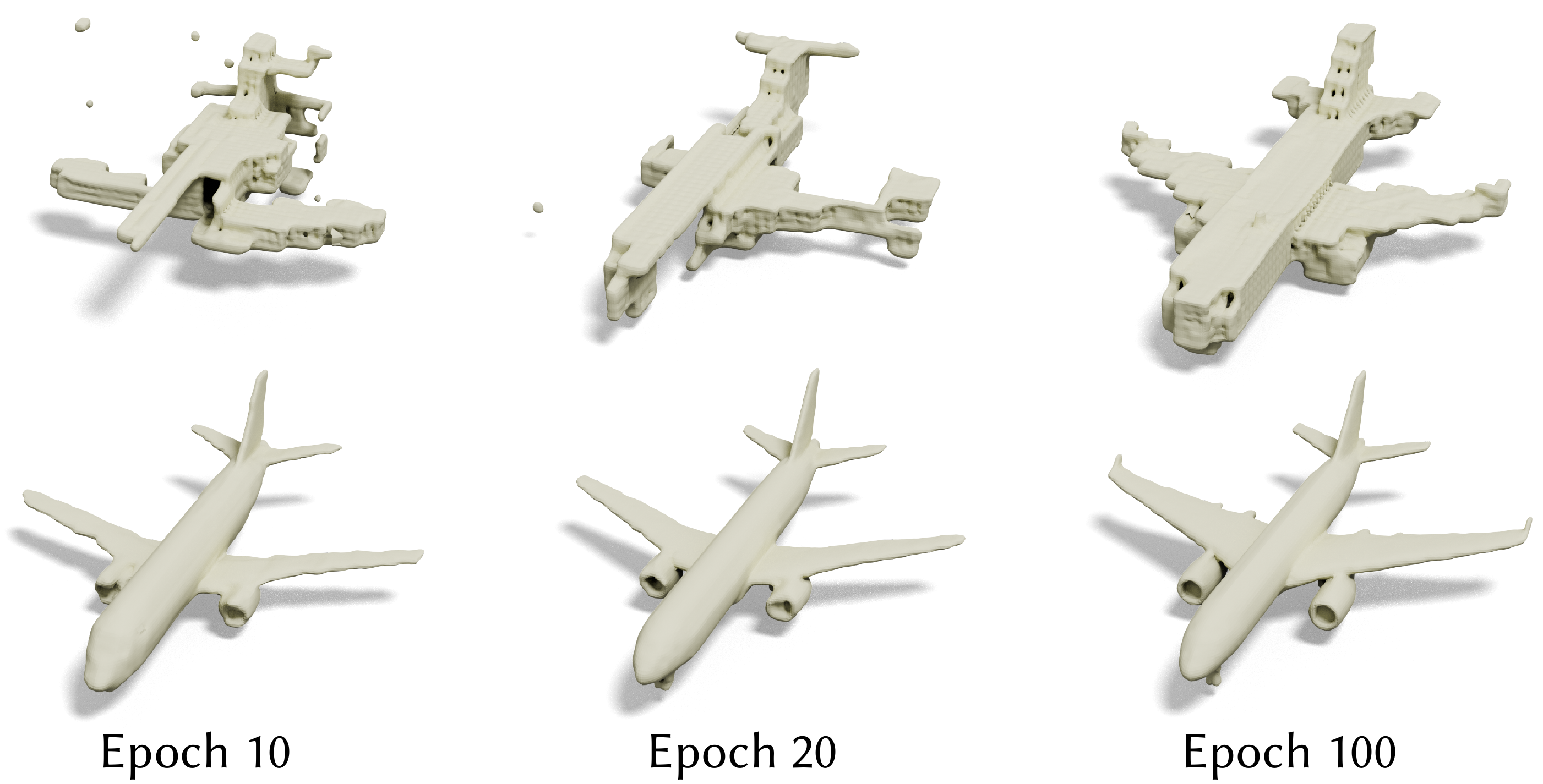}
  \caption{Comparision with 3D causal autoregressive model. Experiments are conducted on the airplane category.}
  \label{fig:ablation-causal}
\end{figure}

\paragraph{Multiscale Autoregressive Models}
One of our key design innovations is training autoregressive models on serialized multiscale binary tokens induced by octrees.
These multiscale binary tokens effectively decompose the challenging task of modeling 3D positions into a series of simpler binary classification tasks, which significantly enhances both training efficiency and the modeling capacity of autoregressive models.
We compare our model with a baseline autoregressive model trained on single-scale 3D positions.
Apart from the multiscale binary tokens, octrees can also be fully determined by their finest nodes, which are described by sparse 3D positions.
We sort these 3D positions in rasterized order, as done in 3DILG~\cite{Zhang2022a} and MeshGPT~\cite{Siddiqui2024}, and train autoregressive models on these positions.
To highlight the effectiveness of our multiscale binary tokens, we maintain identical model architecture and training settings for a fair comparison.
The training results are shown in \cref{fig:ablation-causal}.
With our multiscale binary tokens, our model achieves a significant improvement in FID score from 142.92 to 27.47 and generation quality.
Furthermore, the convergence speed is much faster: after just 10 epochs, our model outperforms the baseline model, which was trained for 100 epochs. \looseness=-1

\begin{figure}
  \centering
  \centering
  \includegraphics[width=0.99\linewidth]{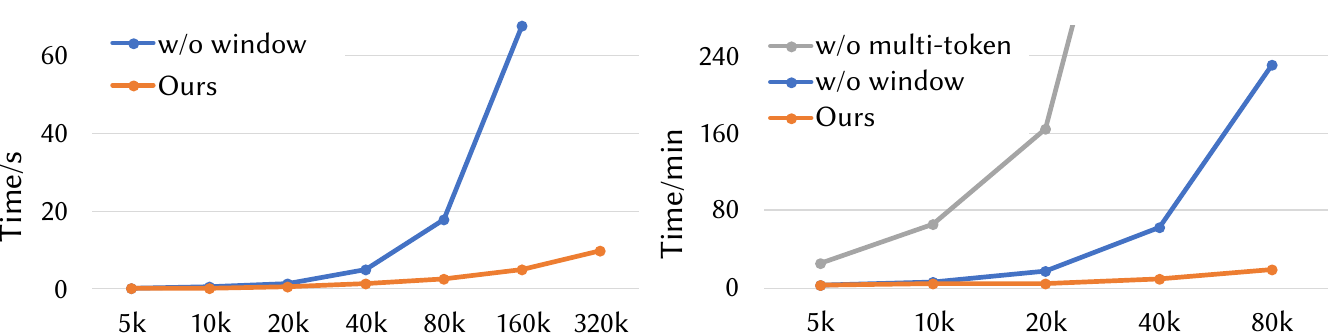}
  \caption{Ablation study on efficiency.
  The time consumption in training and generation are shown in the left and right plots, respectively. The X-axis represents the token length. The training time is measured per iteration, and the generation time is measured per mesh. }
  \label{fig:efficiency}
\end{figure}

\paragraph{Efficiency}
The second key design innovation is extending octree-based window attention to our multiscale binary sequence and introducing multiple token prediction for efficient training and generation.
We analyze the impact of these designs on the efficiency of our model using a single NVIDIA 4090 GPU.
The results are shown in \cref{fig:efficiency}.
It can be seen that we speed up training by $13\times$ when the token length exceeds $160k$ compared to standard global attention, and we improve generation speed by over $69\times$ when the token length exceeds $40k$ compared to single-token prediction.
These designs bring significant efficiency improvements to our model, making our experiments feasible on 4 NVIDIA 4090 GPUs, which allows our model to be easily scaled up and makes it more accessible to a wider range of researchers. \looseness=-1

\paragraph{Position Encoding}
The proposed RoPE3D and learnable depth embeddings are crucial for our model to capture the spatial and scale relationships among tokens.
We conduct experiments to assess the impact of these components by removing them sequentially.
The results are shown in \cref{tab:ablation}. \looseness=-1

\paragraph{Z-order}
Previous 3D autoregressive models rely on rasterization-ordered positions~\cite{Zhang2022a,Siddiqui2024}, whereas we adopt the z-ordering induced by octrees.
By preserving spatial locality, z-ordering allows the transformer to more effectively model highly correlated tokens.
We validate the advantages of z-ordering through ablation studies, as shown in \cref{tab:ablation}. \looseness=-1

\paragraph{Tradeoff between Quality and Speed}
Our model supports parallel token generation, enabling a flexible tradeoff between generation speed and the quality of the generated shapes.
Specifically, we can generate more tokens in parallel to accelerate the generation process, while sacrificing some quality.
In the extreme case, we can generate one token at a time, which corresponds to the standard autoregressive generation approach.
To analyze this tradeoff, we conduct experiments to evaluate the relationship between the number of tokens generated in parallel and the generation speed.
The results are shown in \cref{tab:ablation}. \looseness=-1

\begin{figure}
  \centering
  \includegraphics[width=0.85\linewidth]{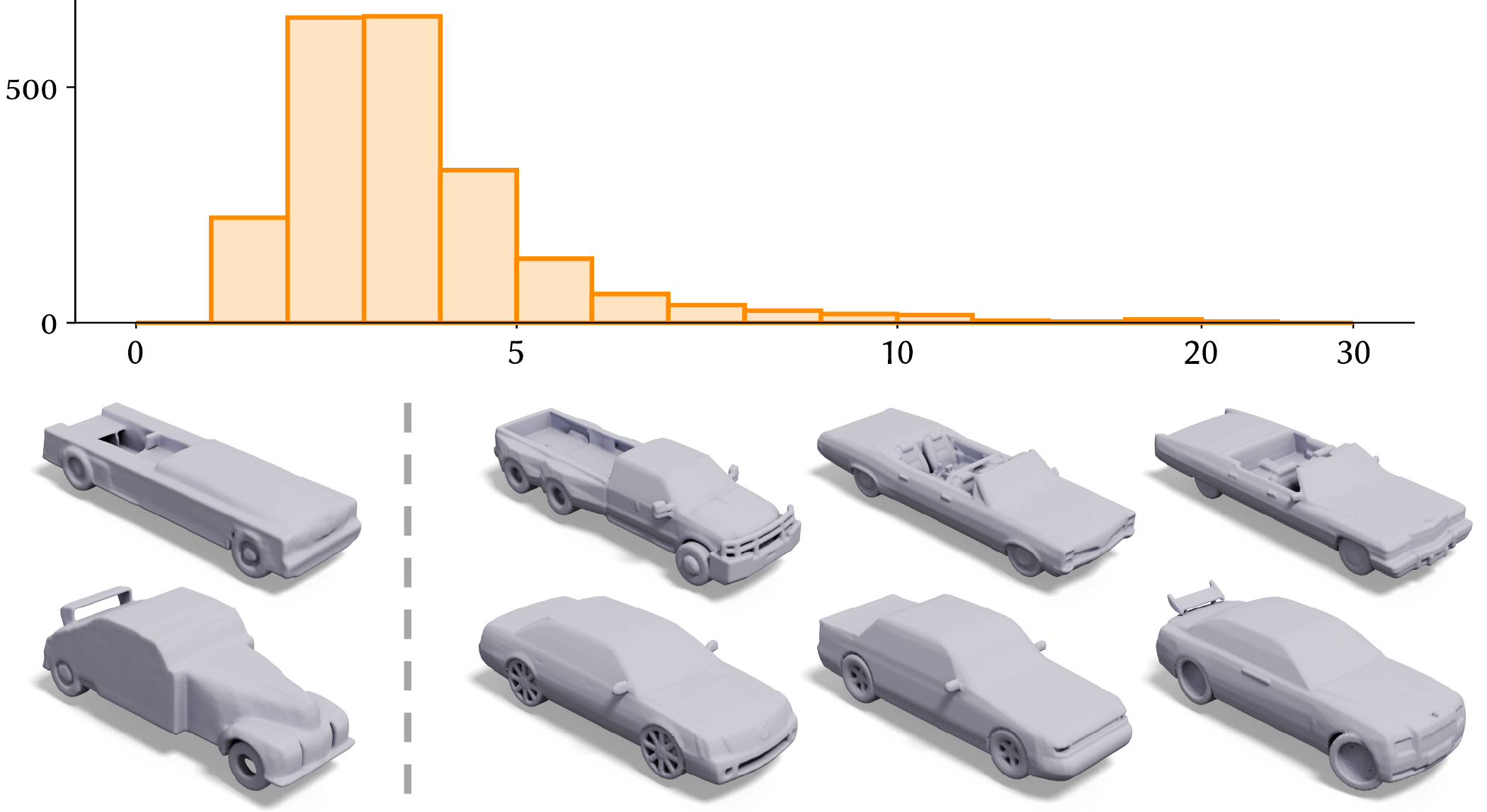}
  \caption{Diversity.
  Top: Histogram of Chamfer Distance (CD) between generated shapes and the training set.
  Bottom: The generated meshes (left) and the 3 nearest shapes (right) from the training dataset according to CD.
  }
  \label{fig:ablation-diversity}
\end{figure}

\begin{table}[t]
  \centering
  \caption{Ablation studies on positional encodings and z-order (left) and number of generation iterations (right).
  The FID values are reported for the airplane category. Time indicates the time consumption to generate a mesh.
  }
  \label{tab:ablation}
  \begin{subtable}{0.48\linewidth}
    \tablestyle{7pt}{1.2}
    \begin{tabular}{lc}
      \toprule
      \textbf{Method} & \textbf{FID} \\
      \midrule
      w/o RoPE3D & 38.03 \\
      w/o Scale Embeddings & 34.41 \\
      w/o z-order & 43.71 \\
      Ours & \textbf{27.47} \\
      \bottomrule
    \end{tabular}
  \end{subtable}
  \hfill
  \begin{subtable}{0.48\linewidth}
    \tablestyle{9pt}{1.2}
    \begin{tabular}{lcc}
      \toprule
      $\mathbf{N}_{iter}$ & \textbf{FID} & \textbf{Time} \\
      \midrule
      64 & 35.95 & 6.43s \\
      128 & 33.60 & 9.65s \\
      256 & 31.90 & 17.05s \\
      512 & 27.47 & 34.51s \\
      \bottomrule
    \end{tabular}
  \end{subtable}
\end{table}

\paragraph{Diversity}
We follow the protocol from LAS-Diffusion~\cite{Zheng2023} and Wavelet-Diffusion~\cite{Hui2022} to assess the diversity of the generated shapes of our OctGPT.
Specifically, we compute the Chamfer distance between the generated shapes and the training set, and construct a histogram with the x-axis representing the Chamfer distance, as shown in \cref{fig:ablation-diversity}.
The results demonstrate that our model is capable of generating novel shapes with high diversity, as the majority of the generated shapes differ from those in the training set.
In the bottom row of \cref{fig:ablation-diversity}, we display some generated samples alongside the three most similar shapes retrieved from the training set.
\looseness=-1

\subsection{Applications} \label{sec:applications}

\begin{figure}
  \centering
  \includegraphics[width=0.99\linewidth]{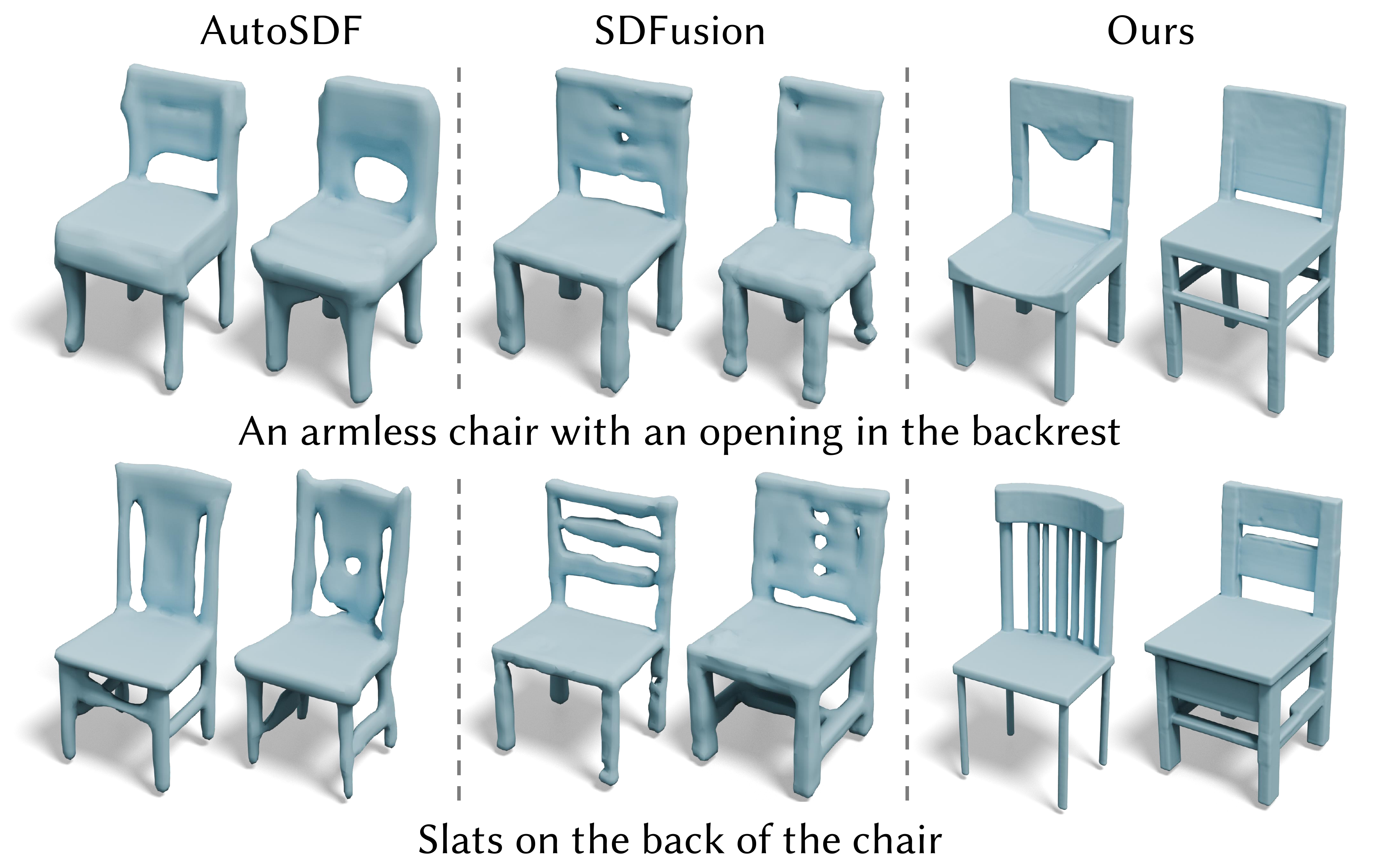}
  \caption{Comparision of text conditional generation results. Experiments are conducted on the chair category of ShapeNet dataset.}
  \label{fig:text}
\end{figure}

\paragraph{Text-Conditioned Generation}
Given text prompts, we leverage text encoder of CLIP~\cite{Radford2021} to extract text features and integrate the extracted features into the generation process through cross-attention modules~\cite{Vaswani2017}.
We compare our method with AutoSDF~\cite{Mittal2022} and SDFusion~\cite{Cheng2023} on Text2Shape dataset~\cite{Chen2018a}, as shown in \cref{fig:text}. We also compare with Shap-E~\cite{Jun2023}, Gaussian Cube~\cite{zhang2024gaussiancube} and LN3Diff~\cite{lan2024ln3diff} on Objaverse dataset, as shown in \cref{fig:objaverse-compare}.
It can be observed that other methods produce shapes that suffer from distortions and artifacts, whereas our method generates shapes with superior quality and greater diversity. \looseness=-1

\begin{figure}
  \centering
  \includegraphics[width=0.99\linewidth]{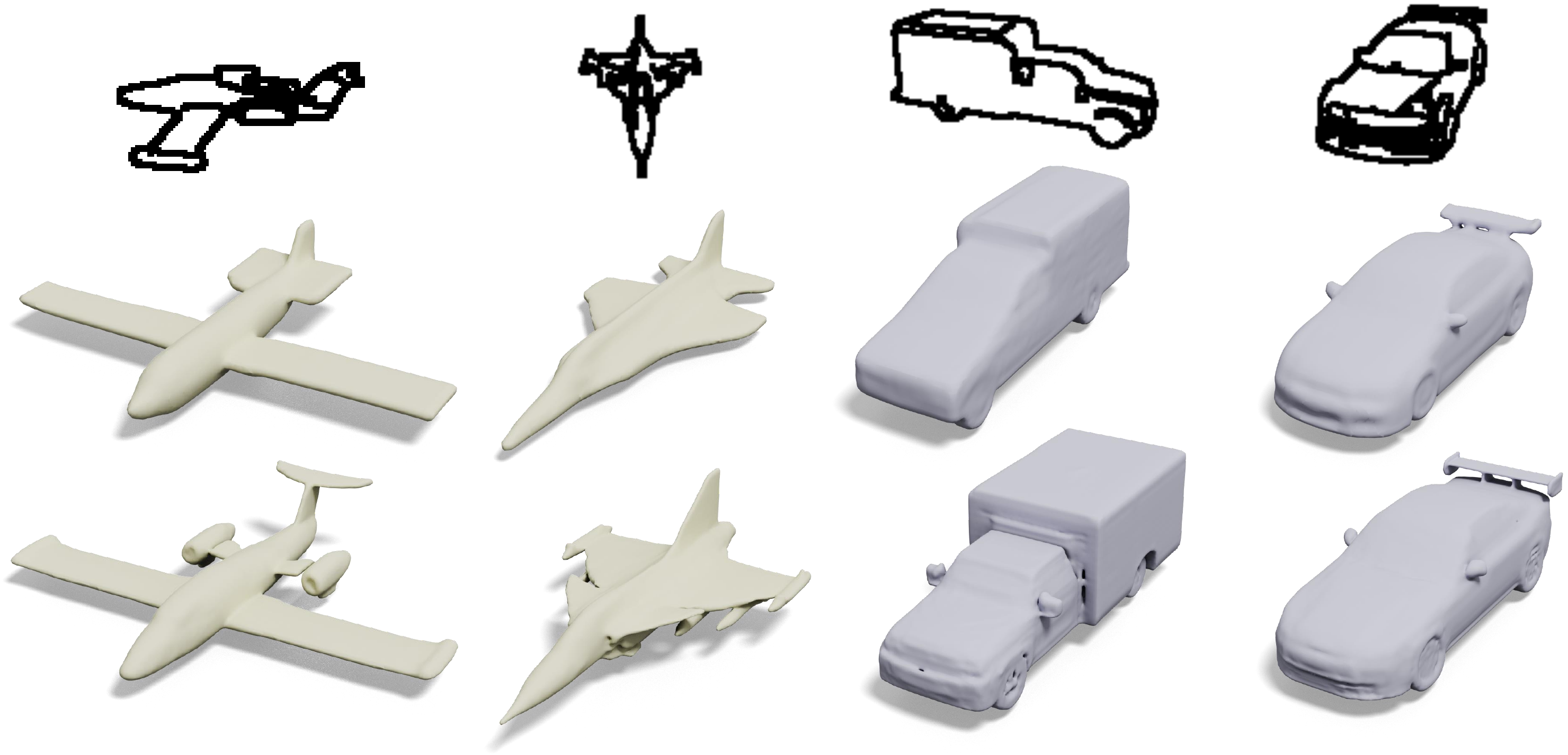}
  \caption{Comparision of sketch conditional generation results. Experiments are conducted on the airplane and car category of ShapeNet dataset. Top: input sketches. Middle row: shapes generated by LAS-Diffusion. Bottom: shapes generated by our OctGPT.}
  \label{fig:sketch}
\end{figure}

\paragraph{Sketch-Conditioned Generation}
We conduct sketch-conditioned generation to demonstrate the generalization ability and versatility of our OctGPT. The sketch data are created following LAS-Diffusion~\cite{Zheng2023}, and we train our model on the airplane and car categories for 100 epochs.
The sketch input is encoded by a pre-trained Dinov2~\cite{Oquab2023} and then integrated into the OctGPT model.
Notably, our model does not rely on view information; instead, we replace the view-aware attention mechanism in LAS-Diffusion with standard cross-attention modules.
Despite this, our model is still able to generate high-quality shapes that are consistent with the input sketches, as shown in \cref{fig:sketch}.
The geometric details are more realistic compared to the results from LAS-Diffusion. \looseness=-1

\begin{figure}
  \centering
  \includegraphics[width=0.99\linewidth]{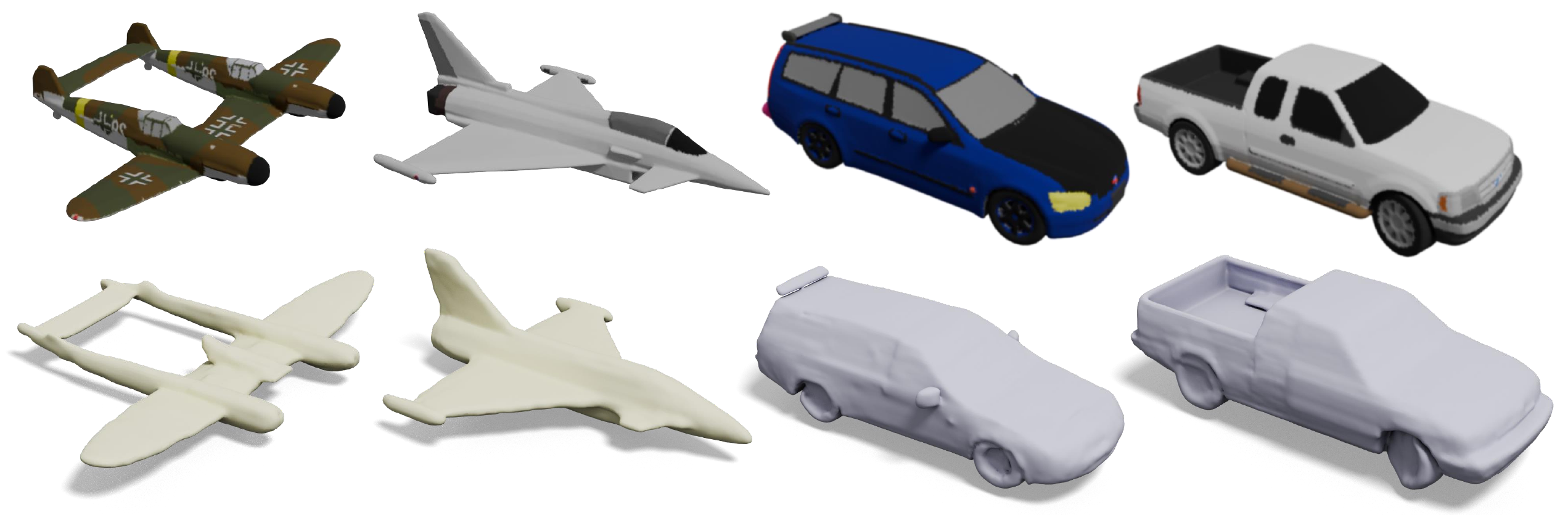}
  \caption{Qualitative results of image conditioned generation. Experiments are conducted on the airplane and car category of ShapeNet dataset.}
  \label{fig:image}
\end{figure}

\paragraph{Image-conditioned Generation}
The sketch-conditioned generation can be easily extended to image-conditioned generation.
Specifically, we utilized rendered images from ShapeNet~\cite{Chang2015} during training. 
The training settings are the same as those used for sketch-conditioned generation.
The qualitative results are shown in \cref{fig:image}.
It can be observed that our method generates shapes that are consistent with the input images and preserve their details.
\looseness=-1

\paragraph{Scene-level Generation}
We also perform a challenging stress test on our OctGPT by generating scenes with multiple objects.
We use the synthetic room dataset~\cite{Peng2020}, which contains $5k$ scenes, each composed by randomly selecting objects from five categories of ShapeNet: chair, sofa, lamp, cabinet, and table.
We train our model on this dataset and generate scenes at a resolution of $1024^3$.
The model is trained for 200 epochs on 4 NVIDIA 4090 GPUs, taking approximately 3 days.
Surprisingly, our model generates diverse scenes with multiple objects, as shown in \cref{fig:scene}.
The generated scenes are visually appealing and exhibit rich details, demonstrating the scalability and versatility of our model.
In comparison, training a standard autoregressive model on the serialized 3D positions results in poor performance, with the model struggling to generate meaningful results due to the complexity of scene-level generation.

\section{Conclusions}

In this paper, we presented \emph{OctGPT}, a novel approach for efficient and high-quality 3D shape generation using a multiscale autoregressive model based on serialized octree representations.
The core idea behind OctGPT is to encode 3D shapes as multiscale binary sequences induced by octrees, and then use an autoregressive transformer model for 3D shape generation. 
We also introduced an efficient transformer architecture that incorporates extended octree attention mechanisms, tailored 3D rotary positional encodings, and parallelized token generation schemes.
OctGPT significantly advances the state-of-the-art in 3D shape generation with autoregressive models,
and scalability, and even surpassing existing diffusion models in some cases.
We demonstrated the scalability and versatility of OctGPT across a wide range of 3D shape generation tasks.

There are a few limitations and exciting future directions.
Firstly, our model is a two-stage pipeline: generating serialized octree representations via VQVAE, followed by 3D shape generation using an autoregressive transformer.
While effective, this approach is not end-to-end trainable, which may constrain the model's overall performance.
Future work can focus on developing end-to-end training strategies. 
Secondly, our model has been trained with limited GPU resources.
We plan to leverage additional computational resources and larger datasets to further enhance scalability and performance.
Lastly, exploring multi-modality training is an exciting direction that could broaden OctGPT’s applicability by enabling it to handle diverse input modalities, which would open up new possibilities for integrating 3D shape generation with other modalities.
 \looseness=-1

\begin{acks}
  This work was supported in part by National Key R\&D Program of China 2022ZD0160801 and Beijing Natural Science Foundation No. 4244081.
  We also thank the anonymous reviewers for their invaluable feedback.
\end{acks}

\bibliographystyle{ACM-Reference-Format}
\bibliography{src/ref/reference}

\vspace{10pt}
\emph{\small{Received January 2025; accepted March 2025; final version April 2025}}

\clearpage
\appendix
\section{Appendix}

\paragraph{Datasets}
\label{sec:dataset}
For comparison, we follow the setup from previous works~\cite{Zheng2023} using five categories from ShapeNet with the same data splitting.
These categories include \textit{chair}, \textit{table}, \textit{airplane}, \textit{car}, and \textit{rifle}.
For meshes in ShapeNet,  we first repair meshes and generate SDFs using the method from~\cite{Wang2022}.
We then sample $100k$ points with oriented normals from the resulting mesh surfaces to construct octrees and $200k$ SDF values for training the VQVAE.
For experiments on Objaverse~\cite{Deitke2023}, we follow the same data processing pipeline, selecting a subset of 45k meshes using the data splitting provided by~\cite{Tang2024}.
Additionally, we use Cap3D~\cite{Luo2023} for text captioning in the dataset.

\paragraph{Implementation Details}
\label{sec:implementation}
To train our VQVAE, we construct octrees with depth 8, corresponding to a resolution of $256^3$.
The encoder downsamples the input octrees to depth 6 and encodes the features into 32-bit binary tokens at the leaf nodes.
The decoder then upsamples the octree back to depth 8.
The AdamW optimizer~\cite{Loshchilov2017} is used with an initial learning rate of $1 \times 10^{-3}$, which linearly decays to $1 \times 10^{-5}$ throughout training.
The VQVAE is trained for 200 epochs on 4 NVIDIA 4090 GPUs, taking approximately 5 days.
For training OctGPT, we use the AdamW optimizer with a constant learning rate of $5 \times 10^{-5}$.
Our model has 170M parameters for ShapeNet dataset, trained on 4 NVIDIA 4090 GPUs for 3 days, and has 440M parameters for Objaverse, trained on 8 NVIDIA 4090 GPUs for 7 days, respectively.
During inference, the number of sampling iterations increases from 64 to 256, while the temperature decreases from 1.5 to 0.5 as the octree depth progresses from depth 3 to 6.
We use the shading-image-based Fr\'echet Inception Distance (FID) to evaluate the quality of generated shapes, as proposed in~\cite{Zheng2023}.
Specifically, we render the generated meshes from 20 pre-determined viewpoints and calculate the FID between the rendered images and the ground-truth images from the training dataset.
A lower FID indicates better quality and diversity of the generated shapes.

\paragraph{More Comparisons}
After VQVAE training, OctGPT adopts an end-to-end architecture for high-resolution generation without relying on cascaded stages or auxiliary models, thereby reducing implementation complexity and mitigating error accumulation.
In contrast, OctFusion requires a two-stage pipeline even after VQVAE training, and XCube relies on a four-stage architecture with significantly more parameters (1.6B) compared to OctGPT (170M).
This unified design allows OctGPT to seamlessly handle complex shape generation. We present comparisons with OctFusion on text-conditioned generation using the Objaverse dataset, as shown in \cref{tab:comparison-text}.
We further compare the efficiency and convergence speed of OctGPT with state-of-the-art autoregressive models, including AutoSDF, 3DILG, and MeshGPT, as illustrated in \cref{tab:comparison-efficiency} and \cref{tab:comparison-convergence}.

\begin{figure*}
    \centering
    \includegraphics[width=1.0\linewidth]{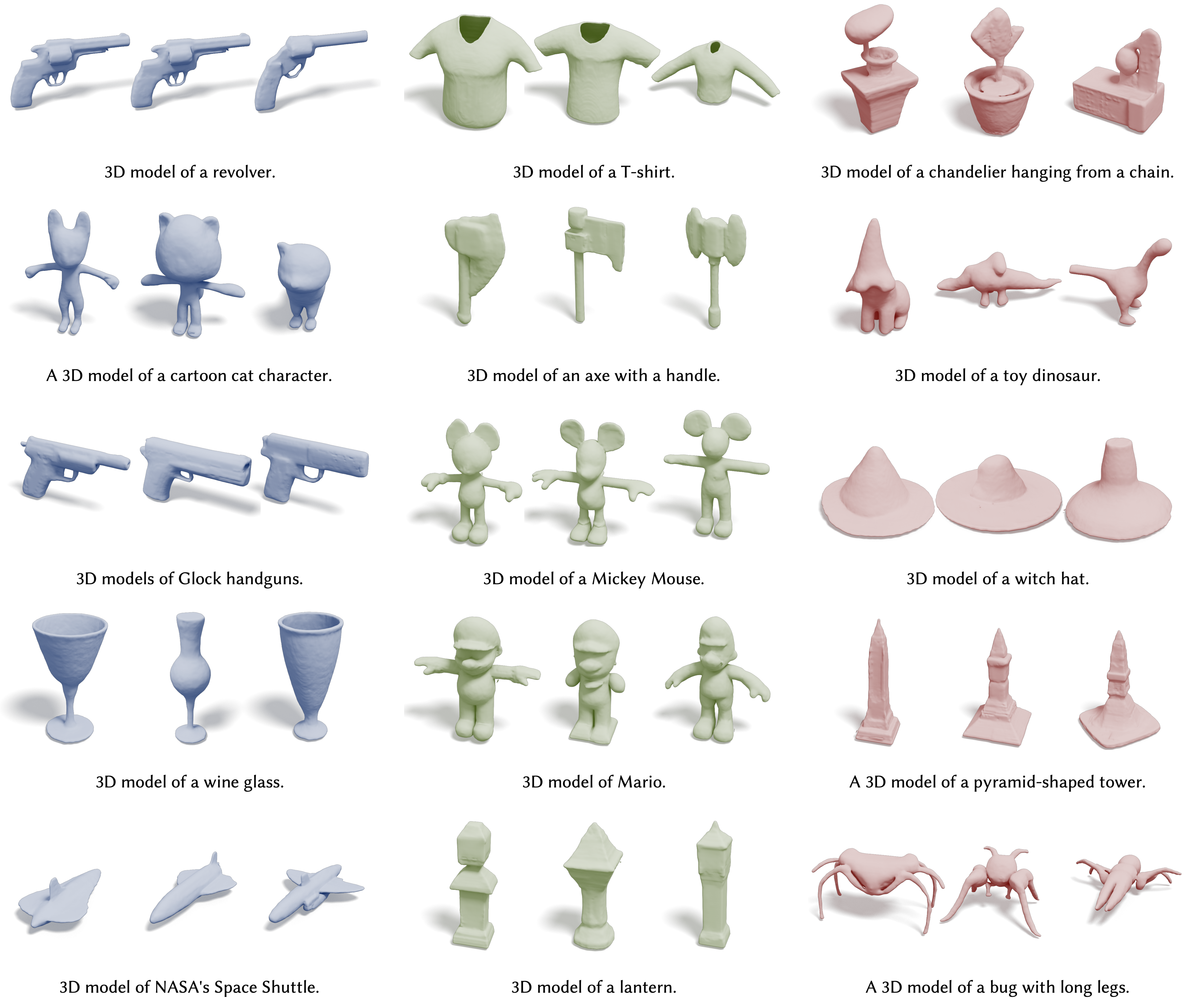}
    \caption{More text-conditioned generation results on Objaverse dataset.}
    \label{fig:objaverse-appendix}
\end{figure*}

\paragraph{More Results}
We present more text-conditioned generation results on Objaverse dataset, as illustrated in \cref{fig:objaverse-appendix}.

\begin{table}[!b]
  \caption{The quantitative comparison of \emph{shading-image-based FID} with OctFusion for text-condition experiment on the Objaverse dataset.}
  \label{tab:comparison-text}
  \tablestyle{20pt}{1.2}
  \begin{tabular}{lll}
    \toprule
    Method & FID (Inception) & FID (CLIP-ViT) \\
    \midrule
    OctFusion & 81.25 & 9.09 \\
    OctGPT & \textbf{59.91} & \textbf{6.45} \\
    \bottomrule
  \end{tabular}
\end{table}

\begin{table}[!b]
  \centering
  \caption{Comparisons of efficiency with state-of-the-art 3D autoregressive models. We compare the training iteration time (in seconds) for longer sequences.}
  \label{tab:comparison-efficiency}
  \tablestyle{9.2pt}{1.2}
  \begin{tabular}{lllllll}
    \toprule
    Method  & 5K  & 10K & 20K & 40K  & 80K  & 160K \\
    \midrule
    AutoSDF & 0.4 & OOM &     &      &      &      \\
    3DILG   & 0.7 & 2.0 & 6.4 & 22.7 & 84.8 & OOM  \\
    MeshGPT & 0.3 & 1.3 & 4.5 & OOM  &      &      \\
    OctGPT  & 0.3 & 0.4 & 0.7 & 1.3  & 2.5  & 5.2  \\
    \bottomrule
  \end{tabular}
\end{table}

\begin{table}[!b]
  \centering
  \caption{Comparisons of convergence speed with state-of-the-art 3D autoregressive models. We present the number of parameters, epochs and GPUs}
  \label{tab:comparison-convergence}
  \tablestyle{17pt}{1.2}
  \begin{tabular}{lllllll|lll}
    \toprule
    Method  & Params & Epochs & GPUs \\
    \midrule
    AutoSDF & 32M & 400 & N/A \\
    3DILG   & 316M & 400 & 4$\times$A100 \\
    MeshGPT & 213M & 2000 & 4$\times$A100 \\
    OctGPT  & 170M & 200 & 4$\times$4090 \\
    \bottomrule
  \end{tabular}
\end{table}

\end{document}